%% file: main.tex
\documentclass[acmsmall,screen,nonacm]{acmart}

\setcopyright{none}
\acmDOI{}
\acmConference{}{}{}
\acmISBN{}

\author{Fares Trad}
\affiliation{%
  \institution{York University}
  \city{Toronto}
  \country{Canada}
}
\email{fares610@my.yorku.ca}

\author{Yizhou Chen}
\affiliation{%
  \institution{Peking University}
  \city{Beijing}
  \country{China}
}
\email{yizhouchen@stu.pku.edu.cn}

\author{Hung Viet Pham}
\affiliation{%
  \institution{York University}
  \city{Toronto}
  \country{Canada}
}
\email{hvpham@yorku.ca}

\author{Zeyu Sun}
\affiliation{%
  \institution{Institute of Software, Chinese Academy of Sciences}
  \city{Beijing}
  \country{China}
}
\email{zeyu.zys@gmail.com}

\author{Gias Uddin}
\affiliation{%
  \institution{York University}
  \city{Toronto}
  \country{Canada}
}
\email{guddin@yorku.ca}

\usepackage{graphicx,booktabs,multirow,multicol}
\usepackage{tabularx}
\usepackage{array}
\usepackage{tcolorbox}
\usepackage{xurl}
\usepackage{tikz}
\usepackage{xcolor}
\usepackage{xspace}

\input{macros}

\usetikzlibrary{arrows.meta, positioning}

\definecolor{darkgreen}{RGB}{34,139,34}

\title{Structured Security Explanations for Learning-Based Vulnerability Detection}

\makeatletter
\def\ps@standardpagestyle{%
  \def\@oddhead{}%
  \def\@evenhead{}%
  \def\@oddfoot{\hfil\thepage\hfil}%
  \def\@evenfoot{\hfil\thepage\hfil}%
}
\makeatother

\begin{document}

\begin{abstract}
\input{sections/00_abstract}
\end{abstract}



\maketitle

\pagestyle{standardpagestyle}
\thispagestyle{standardpagestyle}

\input{sections/01_Introduction}

\input{sections/02_related_work}

\input{sections/03_approach}
\input{sections/04_experimental_setup}
\input{sections/05_empirical_study}
\input{sections/06_discussion}
\input{sections/07_threats_to_validity}
\input{sections/08_conclusion}
\input{sections/09_Data_Availability}

\bibliographystyle{ACM-Reference-Format}
\bibliography{references}

\end{document}

%% file: macros.tex
\newcommand{\tool}{\textsc{VulnLens}\xspace}
\newcommand{\LLM}{\textsc{Explainer}\xspace}
\newcommand{\SM}{\textsc{Detector}\xspace}

%% file: sections/00_abstract.tex
Software vulnerability detection remains difficult because existing approaches either rely on predefined analysis rules, learn vulnerability patterns implicitly from code and labels, or use large language models (LLMs) themselves to produce the final security judgment. These approaches can be limited by rigid analysis rules, implicit learning of security-relevant program behavior, or reliance on the LLM itself for the final decision. This leaves open whether structured security-relevant program behavior can be made explicit and learned jointly with source code for vulnerability detection.

We present \tool{}, an explanation-augmented framework for binary C/C++ vulnerability detection that separates program-behavior analysis from the final vulnerability decision. A label-blind LLM-based \LLM{} analyzes each function from seven fixed program-analysis perspectives, including data flow, memory access, object lifetime, arithmetic and type behavior, control flow, shared state, and external-resource access, without receiving vulnerability labels or CWE information. A supervised \SM{} then combines the source code with these structured explanations through sequential cross-attention to learn vulnerability-relevant code--explanation relationships.

We evaluate \tool{} on a PrimeVul-derived dataset covering ten CWE categories and assess cross-dataset transfer on SecVulEval, a separate benchmark never used for training or retraining. On the 526-function PrimeVul test set, \tool{} achieves 75.5\% accuracy and 77.2\% F1. Among the closest direct-LLM baselines, \tool{} exceeds DeepSeek V4 Pro by 4.2 percentage points in accuracy and Gemini 2.5 Pro by 2.2 percentage points in F1. Compared with a separately trained code-only detector, matched explanations improve F1 by 7.3 percentage points. When transferred directly to SecVulEval without retraining, \tool{} achieves 66.4\% accuracy and 65.8\% F1, outperforming DeepSeek V4 Pro by 7.6 points in accuracy and 20.0 points in F1, and Gemini 2.5 Pro by 16.6 points in accuracy and 7.8 points in F1.

%% file: sections/01_Introduction.tex
\section{Introduction} 
\label{sec:introduction}
\input{figures/01_vulnlens_detector_example}

Software vulnerability detection aims to identify security-relevant flaws in programs that may be exploited to compromise confidentiality, integrity, or availability. Such flaws can have substantial real-world consequences. For example, the 2017 Equifax breach exposed the personal information of approximately 147 million people and resulted in a settlement of at least \$575 million following allegations that Equifax failed to address known security weaknesses in its systems~\cite{ftcEquifax2019}. Despite extensive work on automated techniques, reliably detecting vulnerabilities in source code remains difficult~\cite{chess2004static,chakraborty2020reveal,sun2024llmreview}.

Existing vulnerability-detection approaches broadly follow three technical pathways. First, static analyzers such as Clang Static Analyzer, Semgrep, and Cppcheck use predefined rules and program models to identify potentially unsafe program behavior~\cite{clangStaticAnalyzer,semgrep,cppcheck}. Their reliance on manually constructed rules and analysis models can make them costly to develop and maintain and can limit their flexibility across vulnerability patterns~\cite{charoenwet2024empirical,kaur2020comparative}. Second, learning-based detectors learn vulnerability-related patterns from labeled source code using representations derived from tokens, program structure, or pretrained code models~\cite{li2018vuldeepecker,li2021sysevr,zhou2019devign,fu2022linevul,feng-etal-2020-codebert,guo2021graphcodebert,wang-etal-2021-codet5,guo-etal-2022-unixcoder}. More recently, LLMs have been applied to vulnerability detection through prompting, instruction tuning, reasoning, and additional program context, demonstrating promising but still imperfect detection performance~\cite{du2024vulllm,lu2024grace,lekssays2025llmxcpg}. Despite these advances, each pathway has limitations in how vulnerability-relevant program behavior is represented and used for detection.

These approaches nevertheless expose vulnerability-relevant program behavior in different and incomplete ways. Static analysis makes program properties explicit, but does so through predefined rules and analysis models. Learning-based detectors are more flexible, but primarily rely on source code and vulnerability labels, requiring them to infer vulnerability-relevant semantics and code--vulnerability relationships implicitly from supervision. Without explicit semantic guidance, the detector must learn which program behaviors are relevant to vulnerability detection from the labels alone. LLM-based approaches can provide additional context and intermediate reasoning~\cite{wei2022cot,nye2021scratchpads}; however, it remains unclear whether the generated context consistently captures the program behavior most relevant to vulnerability detection. Moreover, existing LLM-based vulnerability-detection approaches commonly use the LLM itself to produce the final security judgment~\cite{du2024vulllm,lu2024grace,lekssays2025llmxcpg}. These limitations motivate a complementary question: \emph{Can we explicitly provide vulnerability-relevant program knowledge and learn useful relationships between this knowledge and source code for detection?}

We propose \textbf{\tool}, an explanation-augmented framework for binary C/C++ vulnerability detection that explicitly separates program-behavior analysis from the final vulnerability decision. We focus on C/C++ because our evaluation benchmarks consist of C/C++ functions; however, the framework itself does not rely on language-specific analysis rules and can be extended to other programming languages. To make the additional knowledge consistently relevant to vulnerability detection, an LLM-based \LLM{} examines every function from seven fixed vulnerability-related perspectives: data flow, memory access, object lifetime, arithmetic and type behavior, control flow, shared state, and external-resource access. These perspectives determine what program behavior the \LLM{} should describe rather than allowing the added context to be unconstrained.

The resulting human-readable explanations make vulnerability-relevant program semantics explicit, providing semantic guidance beyond the binary vulnerability label. A supervised \SM{}, trained to predict the vulnerability label, then jointly learns from the source code and these explanations through cross-attention, which relates the explanation representations to the source-code representation. The \LLM{} receives neither the ground-truth label nor dataset-assigned CWE and is explicitly prohibited from producing a vulnerability verdict, so its role is to provide behavioral evidence rather than perform the final classification. Figure~\ref{fig:vulnlens_detector_example} illustrates this design using a vulnerable QEMU function from CVE-2019-3812 (CWE-125)~\cite{nvd-cve-2019-3812}. The function reads \texttt{edid\_blob[s->reg]} and then increments \texttt{reg} without a visible bounds check. GPT-5.4, when given only the source code, classified it as non-vulnerable. In contrast, the generated memory-access explanation identifies that no visible check constrains \texttt{s->reg} to the valid range of \texttt{edid\_blob} before the read. \tool correctly classifies the function as vulnerable; the corresponding security patch constrains the index using the size of \texttt{edid\_blob}~\cite{qemu-cve-2019-3812-fix}.

We evaluate \tool on a PrimeVul-derived dataset~\cite{ding2025primevul} covering ten CWE categories. In our manual audit, two reviewers rated an average of 98.6\% of GPT-5.4 explanations as fully correct (97.6\% and 99.5\%, respectively; Section~\ref{sec:rq1}). On the 526-function held-out test set, \tool achieved 75.5\% accuracy and 77.2\% F1. Among the direct-LLM baselines, \tool outperformed DeepSeek V4 Pro by 4.2 percentage points in accuracy, Gemini 2.5 Pro by 2.2 points in F1, and GPT-5.5 by 7.4 points in accuracy and 14.2 points in F1 (Table~\ref{tab:test_set_results}). We further study explanation source, detector backbone, prompt design, input components, failure cases, and cost.

This paper makes the following contributions:

\begin{itemize}
    \item \textbf{Explanation-first framework:} We introduce a vulnerability-detection framework in which a label-blind \LLM{} generates structured program-analysis evidence rather than vulnerability verdicts.
    \item \textbf{Structured perspectives \& cross-attention fusion:} We design seven guided analysis perspectives and a cross-attention \SM{} that learns vulnerability-relevant relationships between source code and generated explanations.
    \item \textbf{Comprehensive empirical evaluation:} We evaluate \tool against direct LLMs and static analyzers, studying transferability, explanation quality, model choices, component contributions, failure cases, and cost.
\end{itemize}

%% file: figures/01_vulnlens_detector_example.tex
\begin{figure}[t]
    \centering
    \includegraphics[width=0.88\linewidth]{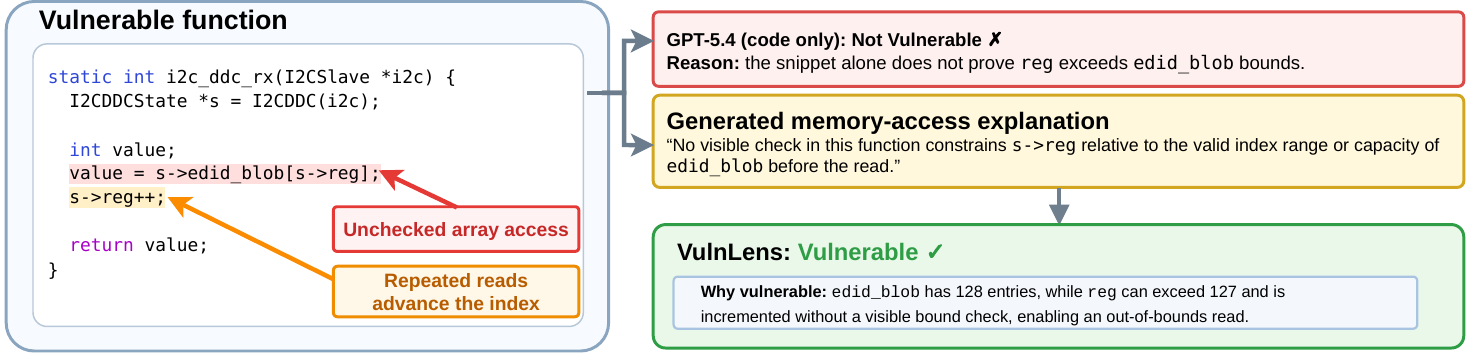}
    \caption{An example of results from different vulnerability detectors.}
    \label{fig:vulnlens_detector_example}
\end{figure}


%% file: sections/02_related_work.tex
\section{Related Work}
\label{sec:related_work}

\noindent{\textbf{Static Vulnerability Detection.}}
Static analysis identifies security weaknesses using techniques such as data-flow analysis, taint tracking, and path-sensitive reasoning~\cite{chess2004static,livshits2005finding,yamaguchi2014modeling}. However, its effectiveness varies across tools, vulnerability patterns, and evaluation settings~\cite{charoenwet2024empirical,kaur2020comparative}. Unlike these approaches, \tool does not rely on predefined analysis rules to produce the final prediction; instead, it learns from source code together with structured behavioral explanations.

\noindent{\textbf{Learning-Based Vulnerability Detection.}}
Learning-based approaches learn vulnerability patterns from code tokens, program slices, and structural representations~\cite{li2018vuldeepecker,li2021sysevr,russell2018automated}. Later systems incorporate code property graphs, graph neural networks, and pretrained transformers to capture program semantics and localize vulnerable statements~\cite{zhou2019devign,chakraborty2020reveal,cheng2021deepwukong,fu2022linevul,li2021ivdetect}. These methods primarily learn from code and labels, without explicit supervision of the security reasoning underlying each label. In contrast, \tool augments source code with structured, label-blind descriptions of program behavior.

\noindent{\textbf{LLM-Based Vulnerability Detection.}}
Recent studies apply LLMs to vulnerability detection and show that performance depends strongly on the prompt, program context, and evaluation setting~\cite{zhou2024llmvd,steenhoek2023empirical,sun2024llmreview}. Proposed approaches enhance detection through retrieved demonstrations, chain-of-thought prompting, instruction tuning, and program-graph augmentation~\cite{lu2024grace,nong2024cot,du2024vulllm,mahyari2024harnessing,lekssays2025llmxcpg}. These methods generally use the LLM as the final detector or rely on its reasoning at inference time. In contrast, \tool separates behavioral analysis from classification: the \LLM generates structured program-analysis explanations, while the \SM learns the final vulnerability decision.

\noindent{\textbf{Explanation-Augmented Learning.}}
Prior vulnerability-detection methods improve interpretability by identifying or highlighting security-relevant statements~\cite{li2021ivdetect,zheng2021vulspg,sejfia2024toward}. More broadly, chain-of-thought and scratchpad prompting demonstrate the value of intermediate reasoning~\cite{wei2022cot,kojima2022zeroshotcot,nye2021scratchpads}, while rationale distillation shows that generated explanations can supervise smaller, specialized models~\cite{wiegreffe2021teach,hsieh2023distilling}. Our work connects these directions by using structured, label-blind program-analysis explanations as inputs to a supervised vulnerability detector rather than as post-hoc descriptions or direct LLM reasoning traces. We evaluate this approach on PrimeVul, a real-world vulnerability dataset designed to address weaknesses in prior data construction~\cite{ding2025primevul}.

%% file: sections/03_approach.tex
\section{Approach}
\label{sec:approach}

\input{figures/03_Overview}

\subsection{Overview}
\label{sec:approach_overview}

This section presents \tool, an explanation-augmented vulnerability detection framework that separates describing program behavior from deciding whether it is vulnerable. Figure~\ref{fig:vulnlens_overview} shows the overall workflow. Given a C/C++ function, \tool produces a binary vulnerability prediction for that function. The framework consists of two modules:

\begin{itemize}
    \item \textbf{\LLM{}} module generates seven structured explanations of the function from program-analysis perspectives, using neither the ground-truth label nor the dataset-assigned CWE, and passes the explanations to the \SM{} module.
    \item \textbf{\SM{}} module encodes the source code and the seven explanations with a shared frozen Transformer, integrates the explanations through sequential cross-attention, and predicts the binary vulnerability label.
\end{itemize}

Together, these modules make behavioral observations available under the same conditions for training and unseen functions and let a separately trained classifier combine those observations with the code.

\subsection{Structured Program-Analysis Explanation Generation}
\label{sec:neutral_explanation_generation}

The goal of this module is to produce, for every function, structured behavioral observations that describe how the code operates without revealing whether it is vulnerable. The module is necessary because a binary label records the outcome of a function but not the behavioral relationships---bounds, object lifetimes, and the checks that guard them---that produce that outcome, and because explanations built from known patches are unavailable for an unseen function. This module therefore derives such observations from the code itself, so that they are available under the same conditions for training and inference.

The input to this module is a single C/C++ function $C$ together with non-label provenance metadata. To cover the behavioral aspects relevant to vulnerabilities while avoiding dependence on any particular defect category, \tool uses seven complementary analysis perspectives:

\begin{enumerate}
    \item data sources and data flow,
    \item memory objects, accesses, and bounds,
    \item object lifetime and ownership,
    \item arithmetic operations and type behavior,
    \item control flow and execution progress,
    \item shared state and concurrent operations, and
    \item path construction and external-resource access.
\end{enumerate}

Table~\ref{tab:neutral-prompt-rationales} shows representative behavioral observations generated from these seven perspectives for the example function in Figure~\ref{fig:vulnlens_detector_example}.

\input{tables/03_app}

Each prompt asks \LLM{} to describe observable program behavior rather than produce a vulnerability verdict. The ground-truth binary label and dataset-assigned CWE are not provided, and the instructions prohibit labels such as \emph{vulnerable}, \emph{non-vulnerable}, \emph{safe}, \emph{unsafe}, \emph{exploitable}, and \emph{non-exploitable}. Claims must be grounded in the visible function or non-label provenance metadata; unavailable information, such as caller constraints or hidden ownership contracts, is recorded as missing context rather than inferred. This label-blind protocol is what makes the explanations usable as ordinary inputs: because they never encode the answer, they can be generated for an unseen function in the same way as for a training function.

Let $C$ denote a function and $\{P_1,\ldots,P_7\}$ the seven prompts. The generation of one explanation per perspective is written as

\begin{equation}
E_i = G(C,m,P_i), \qquad i\in\{1,\ldots,7\},
\label{eq:neutral_explanation}
\end{equation}

where $G$ denotes \LLM{}, $m$ contains non-label provenance metadata such as project and file name, and $E_i$ is the explanation produced under perspective $P_i$. Each $E_i$ is returned as structured JSON containing the perspective, scope, relevant inputs and state, grounded observations, program relationships, missing context, and a concise summary. The schema contains no vulnerability label or confidence field, so the seven explanations can be consumed directly as additional inputs without discarding task-relevant information. In the principal experiment, GPT-5.4 generates the explanations without Semgrep or Cppcheck outputs. The resulting set $\{E_1,\ldots,E_7\}$ is passed to the \SM{} module, which decides whether the function is vulnerable.

\subsection{Semantic Fusion of Code and Explanations}
\label{sec:code_explanation_fusion}

The goal of this module is to learn how the seven explanations relate to the source code and to predict whether the function is vulnerable. The module is necessary because supplying additional analysis at inference time is not sufficient on its own: when an LLM both interprets the analysis and issues the verdict, the resulting decision is not connected to the code by a classifier trained for that purpose. This module instead pairs the explanations with the code in a supervised model, so that the combination is learned against the vulnerability labels.

The input to this module is the source code $C$ together with the seven explanations $\{E_1,\ldots,E_7\}$ produced by the previous module. A shared pretrained Transformer independently encodes both modalities. In the principal experiment, we use \texttt{deepseek-ai/DeepSeek-R1-Distill-Qwen-7B}.

Let $f_{\theta}$ denote the shared encoder:

\begin{align}
H_C^{(0)} &= f_{\theta}(C),\\
H_{E_i} &= f_{\theta}(E_i),
\qquad i\in\{1,\ldots,7\}.
\label{eq:neutral_code_explanation_embeddings}
\end{align}

Here, $H_C^{(0)}$ is the initial code representation and $H_{E_i}$ is the encoding of explanation $E_i$. Reusing one encoder for both modalities keeps the two representations directly comparable. The encoder parameters $\theta$ remain frozen. Training optimizes only the cross-attention, LayerNorm, projection, and classification parameters, which keeps the number of trained parameters small and isolates the effect of the explanations from changes to the code encoder.

The seven explanation channels are fused sequentially. At step $i$, the current code representation is the query and explanation $E_i$ provides the key and value:

\begin{equation}
A_i =
\operatorname{MHA}_i
\left(
H_C^{(i-1)},
H_{E_i},
H_{E_i};
M_{E_i}
\right),
\qquad i\in\{1,\ldots,7\},
\label{eq:neutral_cross_attention}
\end{equation}

Here, $H_C^{(i-1)}$ is the code representation after $i-1$ fusion steps, $M_{E_i}$ is the explanation attention mask, and $A_i$ is the attended explanation signal. The result is incorporated through a residual connection and LayerNorm,

\begin{equation}
\widetilde{H}_C^{(i)}
=
\operatorname{LayerNorm}_i
\left(
H_C^{(i-1)}
+
\operatorname{Dropout}(A_i)
\right),
\label{eq:neutral_residual_update}
\end{equation}

followed by a block-specific projection:

\begin{equation}
H_C^{(i)}
=
\mathbf{W}_i
\widetilde{H}_C^{(i)}
+
\mathbf{b}_i.
\label{eq:neutral_projection}
\end{equation}

Here, $\mathbf{W}_i$ and $\mathbf{b}_i$ are the projection parameters of fusion step $i$, and $H_C^{(i)}$ is the updated code representation. Each explanation position has independent cross-attention, LayerNorm, and projection parameters, so the model can weight each perspective differently instead of treating the seven explanations as a single undifferentiated context. The order is fixed across training and inference: data flow, memory, lifetime, arithmetic, control flow, shared state, and resource access. The final fused representation is $H_C^{(7)}$.

Because sequences differ in length and are padded, the fused representation must be reduced to a fixed dimension without being biased by padding. We obtain a fixed-dimensional representation using mask-aware mean pooling:

\begin{equation}
\operatorname{Pool}(H,M)
=
\frac{
    \sum_{t=1}^{T} M_t H_t
}{
    \max\left(1,\sum_{t=1}^{T} M_t\right)
}.
\label{eq:neutral_masked_pooling}
\end{equation}

where $H$ is a sequence of length $T$ and $M$ is its attention mask, so that only valid positions contribute to the average. The final representation also retains the original code encoding through a residual path:

\begin{equation}
\mathbf{z}
=
\operatorname{Pool}(H_C^{(7)},M_C)
+
\operatorname{Pool}(H_C^{(0)},M_C),
\label{eq:neutral_final_representation}
\end{equation}

where $M_C$ is the source-code attention mask and $\mathbf{z}$ is the final function representation. The residual path keeps the code's own encoding available even when an explanation offers little additional evidence, so the classifier does not have to recover existing code features.

A linear classifier then predicts the binary label:

\begin{align}
\mathbf{o}
&=
\mathbf{W}_{c}\mathbf{z}
+
\mathbf{b}_{c},\\
\widehat{\mathbf{y}}
&=
\operatorname{softmax}(\mathbf{o}),
\label{eq:neutral_classifier}
\end{align}

where $\mathbf{o}\in\mathbb{R}^{2}$ contains the non-vulnerable and vulnerable logits and $\widehat{\mathbf{y}}$ is the predicted label distribution. Training minimizes cross-entropy over the ground-truth binary labels, which is the only place where the labels enter the framework.

%% file: figures/03_Overview.tex
\begin{figure*}[t]
    \centering
    \includegraphics[width=0.92\textwidth]{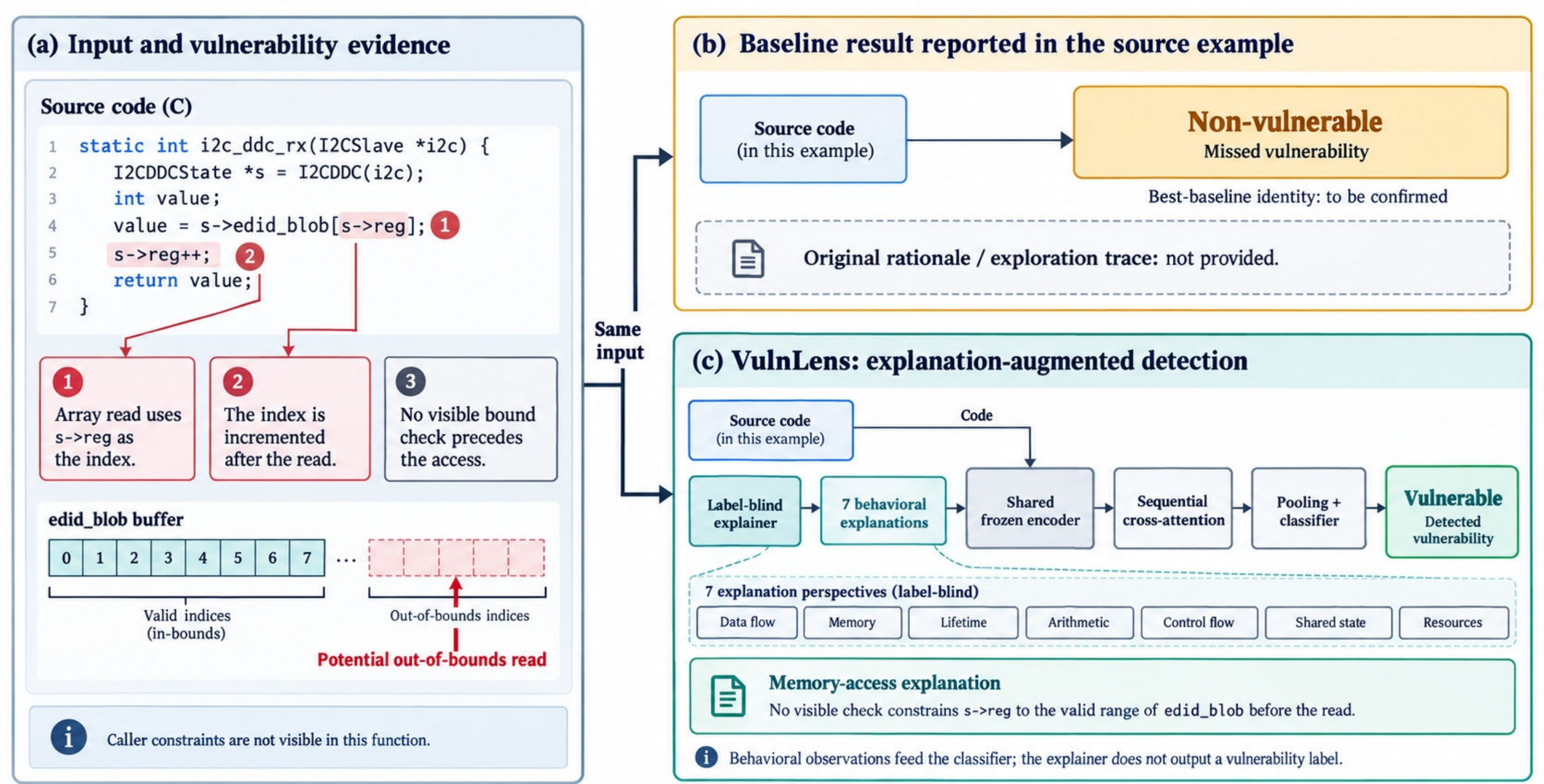}
    \caption{Overview of the VulnLens framework.}
    \label{fig:vulnlens_overview}
\end{figure*}


%% file: tables/03_app.tex
\begin{table*}[t]
\centering
\caption{Example structured explanations for the same input function.}
\label{tab:neutral-prompt-rationales}
\small
\renewcommand{\arraystretch}{1.08}
\begin{tabular}{
    >{\raggedright\arraybackslash}p{0.27\textwidth}
    p{0.65\textwidth}}
\toprule
\textbf{Analysis Perspective} &
\textbf{Representative Generated Observation} \\
\midrule

Data sources and data flow &
The input parameter \texttt{i2c} is used to derive \texttt{s}; state from
\texttt{s->edid\_blob[s->reg]} then flows into \texttt{value} and directly
to the return value. \\

Memory objects, accesses, and bounds &
The function reads \texttt{edid\_blob} at index \texttt{s->reg}. No visible
check constrains \texttt{s->reg} relative to the valid index range or capacity
of \texttt{edid\_blob} before the read. \\

Object lifetime and ownership &
The function derives \texttt{s} from the caller-provided \texttt{i2c} object
and uses its state, but performs no visible allocation, release, or ownership
transfer. \\

Arithmetic operations and type behavior &
The current value of \texttt{s->reg} is used as an index and is then
incremented by one; assigning the selected element to \texttt{int value} may
also involve an implicit conversion depending on the unseen element type. \\

Control flow and execution progress &
Execution follows a single straight-line path: derive \texttt{s}, perform the
indexed read, increment \texttt{s->reg}, and return, with no visible branches,
loops, or alternate exits. \\

Shared state and concurrent operations &
The function reads object state through \texttt{s}, uses \texttt{s->reg} as
an index, and then updates it. No lock, atomic operation, or other
synchronization is visible around this read-then-update sequence. \\

Path construction and external-resource access &
The function constructs no path or external-resource identifier and performs
no visible filesystem, command, or network access; its operations are limited
to the in-memory state referenced through \texttt{s}. \\

\bottomrule
\end{tabular}
\end{table*}

%% file: sections/04_experimental_setup.tex
\section{Experimental Settings}
\label{sec:experimental_settings}

\subsubsection{Dataset}
\label{sec:experimental_dataset}

\input{tables/04_exp_setting}

We evaluate \tool on a C/C++ dataset derived from PrimeVul~\cite{ding2025primevul}. We first identify the ten CWE categories with the largest numbers of functions in PrimeVul: CWE-119, CWE-125, CWE-787, CWE-416, CWE-190, CWE-362, CWE-415, CWE-835, CWE-120, and CWE-22.

For training, we randomly sample 100 functions from each selected CWE within PrimeVul's original training split, yielding 1,000 functions with 476 vulnerable and 524 non-vulnerable samples. For validation and test construction, we retain the vulnerable functions for each selected CWE and randomly sample an equal number of non-vulnerable functions using seed 1337, while preserving each function's original PrimeVul split assignment. This yields 607 validation functions (288 vulnerable, 319 non-vulnerable) and 526 test functions (266 vulnerable, 260 non-vulnerable). The validation set is used for checkpoint selection and the test set for final evaluation.

\subsubsection{Explanation Generation}
\label{sec:experimental_explanation_generation}

We generate seven explanations per function using the perspectives defined in Section~\ref{sec:neutral_explanation_generation}. \LLM receives the function together with non-label provenance metadata, but not the ground-truth label, dataset-assigned CWE, CVE description, commit information, or static-analysis findings.

We compare GPT-5.4 and DeepSeek-R1-Distill-Qwen-32B (DeepSeek-R1-Qwen-32B) as \LLM configurations. GPT-5.4 is accessed through its API, while DeepSeek-R1-Qwen-32B runs locally on the same GPU server used for the remaining experiments. For the seven-prompt configuration, each \LLM generates seven explanations for each of the 2,133 functions across the training, validation, and test sets, yielding 14,931 explanations per \LLM.

For RQ5, we additionally evaluate a CWE-specific-10 prompt set. Unlike seven-prompt, which was designed without assuming knowledge of the target CWE, CWE-specific-10 contains one analysis prompt for each of the ten selected CWE categories. Every function receives all ten prompts, allowing us to evaluate whether explicitly organizing explanations around known candidate weakness categories improves downstream detection.

\subsubsection{Baselines and Ablations}
\label{sec:experimental_baselines}

We compare \tool with three baseline categories and several ablation families covering input components, prompt design, \LLM choice, and \SM backbone choice.

\paragraph{Static-analysis baselines.}
We evaluate Semgrep OSS~v1.86.0 using the Semgrep Registry \texttt{p/default} configuration and Cppcheck~v2.15.0 with all checks enabled. Each function is written to a separate \texttt{.c} or \texttt{.cpp} file according to its original extension. Findings are mapped to functions using the reported file and line locations together with a manifest containing each function's line interval.

We apply a mechanical rule to obtain binary predictions. Semgrep classifies a function as vulnerable when it reports at least one finding with \texttt{WARNING} or \texttt{ERROR} severity. Cppcheck classifies a function as vulnerable when it reports at least one finding with \texttt{warning} or \texttt{error} severity and a non-empty CWE identifier. These severity labels refer to reported findings, not tool execution failures. All other functions are classified as non-vulnerable.

\paragraph{Direct-LLM baselines.}
GPT-5.5, DeepSeek V4 Pro, and Gemini~2.5~Pro receive one request per test function containing the function body and non-label metadata, including sample identifier, source split and line number, project name, and file name. The prompt excludes the ground-truth label, CWE, CVE and commit information, static-analysis findings, and generated explanations. We retain one successful prediction per function and model; requests are repeated only after API errors, malformed responses, or missing binary decisions. 

\paragraph{Explanation baselines.}
GPT-5.4 and DeepSeek-R1-Qwen-32B are each evaluated in direct and self-explanation settings. In the direct setting, the model receives only the code-based classification prompt. In the self-explanation setting, it additionally receives the same seven explanations that it generated for that function and that are used by the corresponding \tool configuration. This tests whether providing the explanatory information directly to the LLM is sufficient without the learned \SM.

\paragraph{Input ablations.}
We compare full \tool with code-only, explanation-removal, and shuffled-explanation variants. The code-only model is trained using source code alone. Explanation removal trains with both modalities but withholds explanations at inference, measuring reliance under a missing-modality distribution shift.

For the shuffled-explanation control, complete seven-explanation bundles are reassigned across training functions using a seeded Sattolo derangement. Perspective identities and bundle structure are preserved, every bundle is used exactly once, and no function receives its own explanations. Source code and labels remain unchanged, while validation and test explanations remain correctly aligned.

\paragraph{Prompt-set ablation.}
We compare seven-prompt and CWE-specific-10 explanations while holding the \LLM, \SM backbone, data, and optimization settings fixed within each configuration.

\paragraph{\LLM ablation.}
We train equivalent \SM configurations using GPT-5.4 or DeepSeek-R1-Qwen-32B explanations with prompts, data, labels, splits, and optimization settings fixed.

\paragraph{\SM backbone ablation.}
We compare frozen DeepSeek-Coder-1.3B, DeepSeek-Coder-6.7B, DeepSeek-R1-Qwen-7B, and Qwen-Coder-7B backbones while holding the \LLM, fusion mechanism, and training data fixed.

\subsubsection{Evaluation Metrics}
\label{sec:experimental_metrics}

We formulate vulnerability detection as binary classification at the function level and evaluate the vulnerable class as positive. Here, true positives (TP) are vulnerable functions correctly predicted as vulnerable, false positives (FP) are non-vulnerable functions incorrectly predicted as vulnerable, true negatives (TN) are non-vulnerable functions correctly predicted as non-vulnerable, and false negatives (FN) are vulnerable functions incorrectly predicted as non-vulnerable. We report accuracy, precision (\(P\)), recall (\(R\)), and F1:

\begin{align}
\mathrm{Precision} &= TP/(TP+FP), &
\mathrm{Recall} &= TP/(TP+FN), \\
\mathrm{F1} &= 2PR/(P+R), &
\mathrm{Accuracy} &= (TP+TN)/(TP+TN+FP+FN).
\end{align}

Precision measures how often a vulnerable prediction is correct, recall measures how many vulnerable functions are detected, and F1 balances the two. Validation F1 is used to select checkpoints.

\tool is implemented in PyTorch and Hugging Face Transformers. The pretrained \SM backbone is frozen; training updates only the cross-attention, LayerNorm, projection, and classification parameters.

Unless otherwise stated, \SM experiments use a maximum sequence length of 2,048 tokens, batch size 1, 32 attention heads per fusion layer, dropout 0.3, AdamW with learning rate $5\times10^{-6}$, unweighted cross-entropy, gradient clipping at 1.0, and bfloat16 weights; each \SM configuration is trained for 50 epochs under five random seeds, and reported results are averaged across the five runs. Within each run, a checkpoint is saved only when validation F1 strictly exceeds the best previously observed value, so the earliest checkpoint is retained in the case of a tie. For each seed, only the checkpoint selected on the 607-function validation set is evaluated on the 526-function test set.

Experiments run primarily on a server with two Intel Xeon Gold 6442Y CPUs, 251~GiB RAM, and four NVIDIA L40S GPUs with approximately 46~GB each. Each training run uses one GPU, while multiple GPUs execute independent experiments concurrently.

\paragraph{Protocol and reproducibility.}
We preserve PrimeVul's original training, validation, and test assignments when constructing our subsets and track sample identifiers to avoid accidental cross-split reuse. The validation split is used only for checkpoint selection, and the test split is used only for the final evaluation reported in Section~\ref{sec:empirical_study}. Test labels are never used during training or checkpoint selection. Across explanation generation and direct-LLM baselines, the models receive no ground-truth vulnerability label, dataset-assigned CWE, CVE description, commit information, or static-analysis findings.

%% file: tables/04_exp_setting.tex
\begin{table}[t]
\centering
\caption{Datasets used in the experiments.}
\label{tab:dataset_splits}
\small
\begin{tabular}{lrrr}
\toprule
\textbf{Split} & \textbf{Functions} & \textbf{Vulnerable} & \textbf{Non-vulnerable} \\
\midrule
Training        & 1,000 & 476 & 524 \\
Validation      & 607   & 288 & 319 \\
Test            & 526   & 266 & 260 \\
\bottomrule
\end{tabular}
\end{table}

%% file: sections/05_empirical_study.tex
\section{Empirical Study}
\label{sec:empirical_study}

We evaluate \tool through nine research questions. RQ1 examines the quality of the explanations that \tool relies on. RQ2 is the main comparison against the baseline categories. RQ3--RQ6 evaluate the design choices of the \SM backbone, the \LLM, the explanation prompts, and the input components. RQ7 examines when \tool succeeds or fails relative to the baselines. RQ8 evaluates transfer to a new dataset, and RQ9 examines the computational cost of the framework.

\begin{itemize}
    \item \textbf{RQ1:} How reliable are the explanations generated by the \LLM?
    \item \textbf{RQ2:} How does \tool{} compare with LLM-based and static-analysis detection?
    \item \textbf{RQ3:} How do \SM backbone architecture and size affect performance?
    \item \textbf{RQ4:} How does the choice of \LLM affect performance?
    \item \textbf{RQ5:} How do different explanation prompt-set configurations affect performance?
    \item \textbf{RQ6:} Which components of \tool contribute most to performance?
    \item \textbf{RQ7:} How does \tool{} succeed or fail relative to direct LLM baselines?
    \item \textbf{RQ8:} How well does \tool{} generalize to an external dataset?
    \item \textbf{RQ9:} What are the training, inference, and cost trade-offs of the approach?
\end{itemize}

\input{sections/rqs/rq1}
\input{sections/rqs/rq2}
\input{sections/rqs/rq3}
\input{sections/rqs/rq4}
\input{sections/rqs/rq5}
\input{sections/rqs/rq6}
\input{sections/rqs/rq7} 
\input{sections/rqs/rq8}
\input{sections/rqs/rq9}

%% file: sections/rqs/rq1.tex
\subsection{RQ1: How reliable are the explanations generated by the \LLM?}
\label{sec:rq1}

We audited a stratified sample of three functions from each of the ten CWE categories. For each
of the resulting 30 functions, we evaluated all seven explanations
generated by GPT-5.4 and DeepSeek-R1-Qwen-32B, yielding 210 explanations
per \LLM and 420 overall. We classified factual quality as
\emph{correct}, \emph{partial}, \emph{incorrect}, or
\emph{hallucinated}, and independently recorded whether each explanation
completely covered the material behavior relevant to its prompt.

The two reviewers rated 205 and 209 of the 210 GPT-5.4 explanations as
fully correct (97.6\% and 99.5\%), respectively; the remaining
explanations were rated partial. For DeepSeek-R1-Qwen-32B, the reviewers
rated 122 and 138 explanations as fully correct and an additional 44 and
57 as partial, respectively. Thus, 166/210 (79.0\%) and 195/210 (92.9\%)
of its explanations were judged either fully correct or partially
correct. A \emph{partial} rating indicates that the main analysis was
supported by the visible code but contained at least one material
inaccuracy or unsupported claim. Completeness was evaluated separately
and captures whether relevant behavior was omitted, rather than whether
the stated analysis was factually correct.

After independent rating and codebook-guided reconsideration of disagreements,
the reviewers agreed on 349 of 420 correctness labels (83.1\%; Cohen's
$\kappa=0.501$) and 356 of 420 completeness judgments (84.8\%;
$\kappa=0.636$). The moderate $\kappa$ values occurred alongside high
raw agreement and a strong prevalence of \emph{correct} judgments, which
increases chance-expected agreement.

\begin{tcolorbox}[boxrule=0.5pt, colback=gray!10, arc=4pt,
left=6pt,right=6pt,top=6pt,bottom=6pt,boxsep=0pt]
\textbf{Answer to RQ1:} GPT-5.4 generated more accurate and complete
explanations than DeepSeek-R1-Qwen-32B, with 83.1\% exact reviewer
agreement on correctness.
\end{tcolorbox}

%% file: sections/rqs/rq2.tex
\subsection{RQ2: How does \tool compare with LLM-based and static-analysis detection?}
\label{sec:rq2}

To answer RQ2, we compare \tool{} with five directly prompted LLMs---GPT-5.4, GPT-5.5, DeepSeek V4 Pro, Gemini~2.5~Pro, and DeepSeek-R1-Qwen-32B---and two static analyzers, Semgrep and Cppcheck, on the 526-function PrimeVul test set. The direct LLM baselines classify functions from code alone; static-analysis predictions are derived from actionable security findings. Table~\ref{tab:test_set_results} reports accuracy, positive-class precision, recall, F1, and false-positive rate (FPR) for these methods and the GPT-5.4--DeepSeek-R1-Qwen-7B configuration of \tool{}.

\input{tables/rq2_1}

\tool{} provides the strongest aggregate accuracy and F1, although different baselines lead on individual metrics. Its accuracy of 75.5\% exceeds DeepSeek V4 Pro, the strongest baseline by accuracy, by 4.2 percentage points. Its F1 of 77.2\% exceeds Gemini, the strongest baseline by F1, by 2.2 points. Relative to Gemini, \tool{} increases precision by 10.9 points and reduces FPR by 28.5 points, while sacrificing 12.8 points of recall. Conversely, its FPR of 31.5\% exceeds DeepSeek V4 Pro's 28.5\%, so the aggregate improvement does not imply uniformly fewer false alarms. Semgrep and Cppcheck achieve precision of 100.0\% and 94.4\%, respectively, but recall of only 0.4\% and 6.4\%, limiting their coverage on this function-level benchmark.

This pattern is consistent with the different decision mechanisms. Direct prompting delegates the final decision to an LLM, whereas \tool{} learns a task-specific decision function from code and structured explanations. The latter can combine explanatory evidence with labeled examples to balance missed vulnerabilities and false alarms. Static analyzers instead rely on the security patterns covered by their configured checks, which can yield precise findings without covering most vulnerable functions in this setting. These aggregate results support the learned classifier's precision--recall balance, rather than superiority on every metric.

\begin{tcolorbox}[boxrule=0.5pt, colback=gray!10, arc=4pt,
left=6pt,right=6pt,top=6pt,bottom=6pt,boxsep=0pt]
\textbf{Answer to RQ2:} \tool{} achieves the highest accuracy and F1 among the evaluated methods, improving on the strongest baseline for each metric by 4.2 and 2.2 percentage points, respectively.
\end{tcolorbox}

%% file: tables/rq2_1.tex
\begin{table*}[t]
\centering
\caption{Overall test-set performance. All metrics are percentages; FPR denotes false-positive rate. Best values are bold.}
\label{tab:test_set_results}
\small
\setlength{\tabcolsep}{5pt}
\begin{tabular}{lccccc}
\toprule
\textbf{Model} &
\textbf{Accuracy} &
\textbf{Precision} &
\textbf{Recall} &
\textbf{F1} &
\textbf{FPR} \\
\midrule

\multicolumn{6}{l}{\textbf{Static analysis}} \\
Semgrep
& 49.6 & \textbf{100.0} & 0.4 & 0.7 & \textbf{0.0} \\
Cppcheck
& 52.5 & 94.4 & 6.4 & 12.0 & 0.4 \\[2pt]
\midrule

\multicolumn{6}{l}{\textbf{Frontier LLMs}} \\
Gemini 2.5 Pro
& 67.9 & 61.9 & \textbf{95.1} & 75.0 & 60.0 \\
GPT-5.5
& 68.1 & 76.1 & 53.8 & 63.0 & 17.3 \\
DeepSeek V4 Pro
& 71.3 & 71.9 & 71.1 & 71.5 & 28.5 \\[2pt]
\midrule

\multicolumn{6}{l}{\textbf{Explanation baselines}} \\
DeepSeek-R1-Qwen-32B (self-expl.)
& 58.4 & 65.2 & 38.0 & 48.0 & 20.8 \\
DeepSeek-R1-Qwen-32B (direct)
& 59.5 & 61.0 & 55.3 & 58.0 & 36.2 \\
GPT-5.4 (self-expl.)
& 65.4 & 82.8 & 39.8 & 53.8 & 8.5 \\
GPT-5.4 (direct)
& 68.3 & 84.6 & 45.5 & 59.2 & 8.5 \\[2pt]

\midrule
\textbf{\tool}
& \textbf{75.5}
& 72.8
& 82.3
& \textbf{77.2}
& 31.5 \\

\bottomrule
\end{tabular}
\end{table*}

%% file: sections/rqs/rq3.tex
\subsection{RQ3: How do \SM architecture and model size affect performance?}
\label{sec:rq3}

To answer RQ3, we compare DeepSeek-Coder-1.3B, DeepSeek-Coder-6.7B, DeepSeek-R1-Qwen-7B, and Qwen-Coder-7B using the same GPT-5.4 seven-prompt explanations, data splits, and hyperparameters. The pretrained encoder remains frozen while the fusion and classification components are trained. Table~\ref{tab:student_backbones} reports test-set accuracy before and after this training. Untrained values summarize five randomly initialized detector heads; trained configurations use the best validation-F1 checkpoint.

\input{tables/rq3_1}

The best overall result comes from DeepSeek-R1-Qwen-7B, but performance does not increase monotonically with parameter count. It achieves 75.5\% accuracy, while DeepSeek-Coder-1.3B reaches 72.1\%, only 3.4 points lower. The 1.3B model also exceeds DeepSeek-Coder-6.7B and Qwen-Coder-7B by 2.3 and 2.5 accuracy points, respectively. Based on the displayed untrained means, training increases accuracy by 21.7, 19.6, 24.7, and 20.8 points for the four backbones, respectively.

The non-monotonic results suggest that the usefulness of frozen representations depends on the backbone's architecture and pretraining as well as its size. In particular, the two 7B Qwen-family backbones differ by 5.9 accuracy points despite similar parameter counts. This comparison does not isolate architecture from pretraining, but it shows that size alone is insufficient to select a detector. The untrained results also exhibit substantial variability because the detector heads are randomly initialized; they measure the effect of learning the downstream components rather than the performance of prompted pretrained models.

\begin{tcolorbox}[boxrule=0.5pt, colback=gray!10, arc=4pt,
left=6pt,right=6pt,top=6pt,bottom=6pt,boxsep=0pt]
\textbf{Answer to RQ3:} Backbone choice affects detection performance, but larger models are not consistently better: DeepSeek-R1-Qwen-7B performs best, while DeepSeek-Coder-1.3B remains within 3.4 accuracy points.
\end{tcolorbox}

%% file: tables/rq3_1.tex
\begin{table*}[t]
\centering
\caption{Detector performance before and after training. Accuracy values are mean $\pm$ standard deviation over five randomly initialized heads; $\Delta$ denotes the improvement in percentage points (pp).}
\label{tab:student_backbones}
\small
\setlength{\tabcolsep}{5pt}
\begin{tabular}{llllrr}
\toprule
\textbf{\SM} &
\textbf{Architecture} &
\textbf{Size} &
\textbf{Version} &
\textbf{Accuracy} &
\textbf{$\Delta$ (pp)} \\
\midrule

\multirow{2}{*}{DeepSeek-Coder-1.3B}
& \multirow{2}{*}{Llama}
& \multirow{2}{*}{1.3B}
& Untrained
& $50.3 \pm 1.5$
& -- \\
&&& Trained
& $72.1 \pm 0.0$
& +21.7 \\
\midrule

\multirow{2}{*}{DeepSeek-Coder-6.7B}
& \multirow{2}{*}{Llama}
& \multirow{2}{*}{6.7B}
& Untrained
& $50.2 \pm 3.8$
& -- \\
&&& Trained
& $69.8 \pm 0.0$
& +19.6 \\
\midrule

\multirow{2}{*}{Qwen-Coder-7B}
& \multirow{2}{*}{Qwen2}
& \multirow{2}{*}{7B}
& Untrained
& $48.8 \pm 2.6$
& -- \\
&&& Trained
& $69.6 \pm 0.0$
& +20.8 \\
\midrule

\multirow{2}{*}{\textbf{DeepSeek-R1-Qwen-7B}}
& \multirow{2}{*}{\textbf{Qwen2}}
& \multirow{2}{*}{\textbf{7B}}
& Untrained
& $50.8 \pm 6.3$
& -- \\
&&& Trained
& $\mathbf{75.5 \pm 0.0}$
& \textbf{+24.7} \\

\bottomrule
\end{tabular}
\end{table*}

%% file: sections/rqs/rq4.tex
\subsection{RQ4: How does the choice of \LLM affect performance?}
\label{sec:rq4}

\input{tables/rq4_1}

RQ4 compares GPT-5.4 and DeepSeek-R1-Qwen-32B as \LLM{}s while
holding the prompts, data, \SM architecture, and training procedure
fixed. GPT-5.4 achieved slightly higher F1 for all three \SM{}s, but
the differences were small: 0.1, 0.6, and 0.8 percentage points for
DeepSeek-Coder-1.3B, DeepSeek-Coder-6.7B, and DeepSeek-R1-Qwen-7B, respectively.

This similarity does not contradict RQ1. RQ1 evaluates whether an
explanation is fully accurate and suitable for a human reader, whereas
downstream performance measures whether it contains signals useful to a
classifier. Many DeepSeek explanations rated as partial still identified
relevant operations and data flows. Moreover, the \SM receives the
source code and seven explanation channels, allowing useful evidence to
remain available even when one explanation contains incomplete or
unsupported details. Thus, explanation quality and predictive utility
are related but not equivalent.

\begin{tcolorbox}[boxrule=0.5pt, colback=gray!10, arc=4pt,
left=6pt,right=6pt,top=6pt,bottom=6pt,boxsep=0pt]
\textbf{Answer to RQ4:} GPT-5.4 produced more accurate human-readable
explanations and consistently achieved higher downstream F1, but
DeepSeek-R1-Qwen-32B remained within 0.8 F1 points across all three \SM{}s.
The \SM was therefore relatively insensitive to the choice of \LLM
in these experiments.
\end{tcolorbox}

%% file: tables/rq4_1.tex
\begin{table}[t]
\centering
\caption{Effect of Explainer choice on Detector performance. All configurations use the same seven prompts, data, and hyperparameters.}
\label{tab:teacher_model_choice}
\small
\setlength{\tabcolsep}{4pt}
\begin{tabular}{llcccc}
\toprule
\textbf{\LLM} &
\textbf{\SM} &
\textbf{Acc.} &
\textbf{Prec.} &
\textbf{Rec.} &
\textbf{F1} \\
\midrule

GPT-5.4
& DeepSeek-Coder-1.3B
& 72.1
& 66.9
& 88.7
& 76.3 \\

DeepSeek-R1-Qwen-32B
& DeepSeek-Coder-1.3B
& 72.4
& 67.6
& 87.2
& 76.2 \\

\midrule

GPT-5.4
& DeepSeek-Coder-6.7B
& 69.8
& 63.5
& 94.7
& 76.0 \\

DeepSeek-R1-Qwen-32B
& DeepSeek-Coder-6.7B
& 70.3
& 64.9
& 89.8
& 75.4 \\

\midrule

GPT-5.4
& DeepSeek-R1-Qwen-7B
& 75.5
& 72.8
& 82.3
& 77.2 \\

DeepSeek-R1-Qwen-32B
& DeepSeek-R1-Qwen-7B
& 74.9
& 72.8
& 80.5
& 76.4 \\

\bottomrule
\end{tabular}
\end{table}

%% file: sections/rqs/rq5.tex
\subsection{RQ5: How do different explanation prompt-set configurations affect performance?}
\label{sec:rq5}

\input{tables/rq5}

Our main configuration uses seven broad program-analysis perspectives rather than prompts tied to specific vulnerability categories. RQ5 examines whether this category-agnostic organization is preferable to a more vulnerability-specific alternative. To answer RQ5, we compare the seven-prompt and CWE-specific-10 explanation sets in two matched \LLM--\SM configurations: GPT-5.4 with DeepSeek-Coder-6.7B, and DeepSeek-R1-Qwen-32B with DeepSeek-R1-Qwen-7B. Within each pair, the \LLM, \SM backbone, training data, evaluation split, and hyperparameters are fixed. Table~\ref{tab:prompt_set_comparison} reports accuracy, precision, recall, and F1. Because the two configurations differ in both prompt organization and number of explanation channels, the comparison evaluates the prompt sets as complete configurations rather than isolating either factor independently.

The seven-prompt configuration yields higher F1 in both matched comparisons despite using fewer explanation channels. With GPT-5.4 and DeepSeek-Coder-6.7B, it improves F1 from 74.2\% to 76.0\%, a gain of 1.8 percentage points. This gain accompanies a 7.1-point increase in recall and a 0.9-point decrease in precision. With DeepSeek-R1-Qwen-32B and DeepSeek-R1-Qwen-7B, it improves F1 by 2.1 points and accuracy by 2.8 points, with increases in both precision (3.3 points) and recall (0.8 points).

A plausible explanation is that the seven prompts describe broadly applicable program properties, such as data flow, memory access, and object lifetime, without asking every function to supply evidence for a predetermined vulnerability category. Such descriptions can provide relevant features even when a particular CWE does not apply. However, because channel count also changes from ten to seven, these results cannot attribute the gain solely to prompt organization or rule out an effect of the amount of explanatory input. They support the seven-prompt configuration as a whole in the two evaluated settings.

\begin{tcolorbox}[boxrule=0.5pt, colback=gray!10, arc=4pt,
left=6pt,right=6pt,top=6pt,bottom=6pt,boxsep=0pt]
\textbf{Answer to RQ5:} The seven-prompt configuration outperforms CWE-specific-10 in both matched configurations, improving F1 by 1.8--2.1 percentage points with fewer explanations.
\end{tcolorbox}

%% file: tables/rq5.tex
\begin{table}[t]
\centering
\caption{7-prompt versus CWE-specific-10 explanations for matched Explainer–Detector configurations. All metrics are percentages.}
\label{tab:prompt_set_comparison}
\small
\setlength{\tabcolsep}{4pt}
\begin{tabular}{lllrrrr}
\toprule
\textbf{\LLM} &
\textbf{\SM} &
\textbf{Prompt set} &
\textbf{Acc.} &
\textbf{Prec.} &
\textbf{Rec.} &
\textbf{F1} \\
\midrule

GPT-5.4
& DeepSeek-Coder-6.7B
& CWE-specific-10
& 69.2 & 64.4 & 87.6 & 74.2 \\

GPT-5.4
& DeepSeek-Coder-6.7B
& 7-prompt
& 69.8 & 63.5 & 94.7 & 76.0 \\

\midrule

DeepSeek-R1-Qwen-32B
& DeepSeek-R1-Qwen-7B
& CWE-specific-10
& 72.1 & 69.5 & 79.7 & 74.3 \\

DeepSeek-R1-Qwen-32B
& DeepSeek-R1-Qwen-7B
& 7-prompt
& 74.9 & 72.8 & 80.5 & 76.4 \\

\bottomrule
\end{tabular}
\end{table}

%% file: sections/rqs/rq6.tex
\subsection{RQ6: Which input components contribute most to performance?}
\label{sec:rq6}

To answer RQ6, we examine the contributions of source code, generated explanations, and their correspondence using DeepSeek-R1-Qwen-7B, the same 1,000-function training set, and the original 526-function test set. We compare full \tool{} with three controls: code-only training, joint training with explanations withheld at inference, and training with explanation sets shuffled across functions. Table~\ref{tab:rq6_modality_ablation} reports accuracy, precision, recall, and F1 for these configurations. The code-only variant assesses the contribution of explanations relative to source code alone, while explanation removal and shuffling examine reliance on explanations and the importance of matching them to the source function, respectively.

\input{tables/rq6_1}

The full model achieves the highest F1, outperforming code-only training by 7.3 percentage points. Adding matched explanations increases accuracy from 67.9\% to 75.5\%, F1 from 69.9\% to 77.2\%, precision from 66.4\% to 72.8\%, and recall from 73.7\% to 82.3\%.

Withholding explanations at inference reduces F1 by 10.1 points to 67.1\%, while training with shuffled explanation sets reduces F1 by 10.4 points to 66.8\%. These controls distinguish complementary information from learned reliance. Removing explanations after joint training introduces a missing-modality distribution shift, so its performance drop measures dependence on explanation inputs rather than their incremental benefit. Shuffling instead disrupts the correspondence between code and explanations. The cleaner contribution estimate comes from the separately trained code-only control: matched explanations improve F1 by 7.3 points and accuracy by 7.6 points, providing evidence that the explanation representation contributes complementary predictive information beyond source code alone.

\begin{tcolorbox}[boxrule=0.5pt, colback=gray!10, arc=4pt,
left=6pt,right=6pt,top=6pt,bottom=6pt,boxsep=0pt]
\textbf{Answer to RQ6:} Full \tool{} improves F1 over code-only training by 7.3 percentage points, while removing or shuffling explanations reduces F1 by 10.1 and 10.4 points, respectively.
\end{tcolorbox}

%% file: tables/rq6_1.tex
\begin{table}[t]
\centering
\caption{Input and explanation ablations using DeepSeek-R1-Qwen-7B. All metrics are percentages.}
\label{tab:rq6_modality_ablation}
\small
\setlength{\tabcolsep}{4pt}
\begin{tabular}{lrrrr}
\toprule
\textbf{Configuration} &
\textbf{Acc.} &
\textbf{Prec.} &
\textbf{Rec.} &
\textbf{F1} \\
\midrule

\textbf{Full \tool}
& \textbf{75.5} & 72.8 & 82.3 & \textbf{77.2} \\

Code only
& 67.9 & 66.4 & 73.7 & 69.9 \\

No explanations at test
& 71.3 & \textbf{79.8} & 57.9 & 67.1 \\

Shuffled training explanation sets
& 62.0 & 59.8 & 75.6 & 66.8 \\

\bottomrule
\end{tabular}
\end{table}

%% file: sections/rqs/rq7.tex
\subsection{RQ7: How does \tool{} succeed or fail relative to direct LLM baselines?}
\label{sec:rq7}

To answer RQ7, we compare the correctness of \tool{} and three direct-LLM baselines---GPT-5.5, DeepSeek V4 Pro, and Gemini~2.5~Pro---on all 526 test functions. For \tool{}, we use predictions from the run with the highest validation F1. We examine agreement patterns and representative cases of shared success, exclusive success by \tool{}, failure by \tool{} despite correct LLM predictions, and shared failure. This analysis characterizes complementary errors in the selected methods; it does not include every baseline evaluated in RQ2.

The methods share many outcomes but also make complementary errors. All four are correct on 199 functions (37.8\%) and incorrect on 25 (4.8\%). \tool{} is the only correct method on 16 functions (3.0\%); 11 of these are non-vulnerable, showing that some exclusive successes arise from avoiding false alarms shared by all three LLMs. Conversely, \tool{} is incorrect on 69 functions (13.1\%) for which at least two LLMs are correct. Thus, the aggregate advantage of \tool{} coexists with complementary errors in which direct LLM classification succeeds.

The selected cases illustrate three forms of disagreement. \tool{} correctly classifies \texttt{r\_\allowbreak bin\_\allowbreak java\_\allowbreak print\_\allowbreak float\_\allowbreak cp\_\allowbreak summary} as non-vulnerable, while all three LLMs infer a vulnerability from an unverified nested-pointer precondition. This case is consistent with a learned detector avoiding a speculative report. All methods correctly identify unsafe length arithmetic in \texttt{string\_number\_format}, but \tool{} misses \texttt{pjmedia\_rtcp\_xr\_rx\_rtcp\_xr}, where the LLMs identify missing validation of variable-length report blocks. Thus, useful explanatory evidence need not translate into a correct detector decision. All methods also miss \texttt{seq\_buf\_alloc}, a thin allocation wrapper whose vulnerability label is not explained by the visible function alone, illustrating the limits of function-level input when relevant behavior depends on callers or callees. These examples illustrate failure mechanisms without establishing their prevalence across all errors.

\begin{tcolorbox}[boxrule=0.5pt, colback=gray!10, arc=4pt,
left=6pt,right=6pt,top=6pt,bottom=6pt,boxsep=0pt]
\textbf{Answer to RQ7:} \tool{} shows complementary errors relative to direct LLMs: it can avoid speculative false positives shared by those baselines, but it can also miss structural validation flaws they detect, while all methods remain limited when relevant security evidence lies outside the provided function.
\end{tcolorbox}

%% file: sections/rqs/rq8.tex
\subsection{RQ8: How well does \tool{} generalize to an external dataset?}
\label{sec:rq-transferability}

\input{tables/rq8_1}

\noindent\textbf{Motivation.}
Performance on PrimeVul may reflect dataset-specific characteristics.
We therefore test whether \tool{} transfers to independently collected
functions without additional training.

Experimental setting. We construct a balanced SecVulEval~\cite{ahmed2026secvuleval} test set of 500 C/C++ functions from the same ten CWE categories, with 25 vulnerable and 25 non-vulnerable functions per CWE. We exclude exact function-code overlaps with all PrimeVul splits.

We evaluate the primary \tool{} configuration---GPT-5.4 seven-prompt explanations with the DeepSeek-R1-Qwen-7B detector---using the RQ7 checkpoint without retraining. We compare it with DeepSeek V4 Pro and Gemini 2.5 Pro, the highest-accuracy and highest-F1 direct-LLM baselines in Section~\ref{sec:rq2}.

\noindent\textbf{Results.}
As shown in Table~\ref{tab:vulnlens_deepseek_v4}, \tool{} achieved
66.4\% accuracy and 65.8\% F1. It outperformed DeepSeek V4 Pro by
7.6 points in accuracy and 20.0 points in F1, primarily through higher
recall (64.8\% vs.\ 34.8\%), although with a higher FPR
(32.0\% vs.\ 17.2\%). Compared with Gemini 2.5 Pro, \tool{} improved
accuracy by 16.6 points and F1 by 7.8 points, with higher precision,
lower FPR, and slightly lower recall.

\begin{tcolorbox}[boxrule=0.5pt, colback=gray!10, arc=4pt,
left=6pt,right=6pt,top=6pt,bottom=6pt,boxsep=0pt]
\textbf{Answer to RQ8:} Without retraining on SecVulEval, \tool{} achieved
the highest accuracy and F1 among the evaluated methods, exceeding
DeepSeek V4 Pro by 7.6 and 20.0 points, and Gemini 2.5 Pro by 16.6
and 7.8 points, respectively.
\end{tcolorbox}

%% file: tables/rq8_1.tex
\begin{table}[t]
\centering
\caption{Transfer performance of VulnLens, DeepSeek V4 Pro, and Gemini 2.5 Pro on SecVulEval.}
\label{tab:vulnlens_deepseek_v4}
\begin{tabular}{lrrrrr}
\toprule
\textbf{Model} & \textbf{Accuracy} & \textbf{Precision} & \textbf{Recall} & \textbf{F1} & \textbf{FPR} \\
\midrule
Gemini 2.5 Pro  & 49.8\% & 49.9\% & 69.2\% & 58.0\% & 69.6\% \\
DeepSeek V4 Pro & 58.8\% & 66.9\% & 34.8\% & 45.8\% & 17.2\% \\
\tool           & 66.4\% & 66.9\% & 64.8\% & 65.8\% & 32.0\% \\
\bottomrule
\end{tabular}
\end{table}

%% file: sections/rqs/rq9.tex
\subsection{RQ9: What are the training, inference, and cost trade-offs of the approach?}
\label{sec:rq9}

\input{tables/rq9_cost_1}

To answer RQ9, we examine explanation generation, Detector training, and inference. Table~\ref{tab:explanation_generation_cost} reports explanation-generation costs for the 1,000-function training set. GPT-5.4 costs \$262.30 in total, or \$0.2623 per function. DeepSeek-R1-Qwen-32B was executed locally; comparable hosted API pricing yields an estimated \$38.15 total, or \$0.0381 per function.

On one NVIDIA L40S, DeepSeek-R1-Qwen-32B generates all seven explanations for a function in approximately 6.27 minutes, corresponding to about 104.5 GPU-hours for the full training set. Detector training requires 0.652 GPU-hours per epoch and 21.69 GiB peak GPU memory.

With explanations cached, Detector inference averages 1.391 seconds per function, with 1.423-second p95 latency and 15.40 GiB peak GPU memory. Explanation generation is therefore a reusable preprocessing cost for cached functions. Previously unseen functions still require seven explanations before Detector inference, so end-to-end latency also depends on the selected Explainer.

\begin{tcolorbox}[boxrule=0.5pt, colback=gray!10, arc=4pt,
left=6pt,right=6pt,top=6pt,bottom=6pt,boxsep=0pt]
\textbf{Answer to RQ9:} Explanation generation is the main upfront
cost: GPT-5.4 costs \$0.2623 per function, while local
DeepSeek-R1-Qwen-32B generation requires approximately 6.27 minutes
per function on one L40S. Detector training requires 0.652 GPU-hours
per epoch and 21.69 GiB peak GPU memory. With explanations cached,
inference takes 1.391 seconds per function with 15.40 GiB peak GPU
memory.
\end{tcolorbox}

%% file: tables/rq9_cost_1.tex
\begin{table}[t]
\centering
\caption{Explanation-generation cost for the 1,000-function training set.
Token counts and total cost aggregate all seven explanations; \$/Expl. and
\$/Func. are averages.}
\label{tab:explanation_generation_cost}
\footnotesize
\setlength{\tabcolsep}{4pt}
\begin{tabular}{lrrrrr}
\toprule
\textbf{\LLM} &
\textbf{Input Tok. (M)} &
\textbf{Output Tok. (M)} &
\textbf{Cost (\$)} &
\textbf{\$/Expl.} &
\textbf{\$/Func.} \\
\midrule
GPT-5.4
& 9.77
& 15.86
& 262.30
& 0.0375
& 0.2623 \\

DeepSeek-R1-Qwen-32B
& 10.07
& 6.79
& 38.15
& 0.0054
& 0.0381 \\
\bottomrule
\end{tabular}

\vspace{2pt}
\parbox{\linewidth}{\footnotesize
DeepSeek was executed locally; the reported monetary cost applies Cloudflare Workers AI rates to the recorded token usage and is a hosted-equivalent estimate.}
\end{table}

%% file: sections/06_discussion.tex
\section{Discussion}
\label{sec:discussion}
\input{figures/06_discussion_model_disagreements}

\subsection{What Do the Explanations Add?}

Full \tool{} substantially outperforms the independently trained code-only detector, achieving 77.2\% versus 69.9\% F1, a gain of 7.3 percentage points. This result suggests that matched explanations provide complementary predictive information beyond source code alone. Removing explanations from the trained fusion model at inference reduces F1 from 77.2\% to 67.1\%, showing substantial reliance on this input. Because this also introduces a missing-modality distribution shift, we treat it as a reliance test rather than an independent estimate of explanation benefit.

Figure~\ref{fig:discussion_model_disagreements} shows one such case. The TensorFlow output is allocated using the shape of \texttt{in0}, while the final operation consumes both \texttt{in0} and \texttt{in1}. The explanations connect these facts and note the absence of a visible compatibility check. \tool{} predicts vulnerable, whereas the code-only detector, the same fusion model without explanations, GPT-5.4, GPT-5.5, and DeepSeek V4 Pro predict non-vulnerable. This case is consistent with explanations helping \SM{} represent cross-object relationships rather than directly supplying the label.

\subsection{Different Models Require Different Evidence}

Direct LLMs differ in how they handle missing program context. Gemini more readily acts on plausible local evidence, while GPT-5.4 and GPT-5.5 more often withhold a vulnerable prediction when the conclusion depends on missing callees, caller contracts, or framework semantics. These differences help explain their precision--recall trade-offs.

\paragraph{DeepSeek V4 Pro.}
Figure~\ref{fig:discussion_model_disagreements} shows a case where DeepSeek V4 Pro alone matches the non-vulnerable label. The other direct LLMs and \tool{} predict vulnerable based on a possible out-of-bounds access after advancing \texttt{cp}. DeepSeek instead notes that confirming the flaw requires unavailable helper semantics. We therefore treat this as a context-ambiguity case rather than evidence that the other predictions are necessarily wrong.

\paragraph{Gemini.}
Figure~\ref{fig:discussion_model_disagreements} shows one of 12 vulnerable samples detected only by Gemini. It identifies that \texttt{input(0)} is accessed at dimension 1 without a visible rank check, matching the security-fixing commit. \tool{} and the other direct LLMs predict non-vulnerable. Gemini's willingness to act on missing preconditions can recover such vulnerabilities, but also contributes to its higher false-positive rate.

\paragraph{GPT-5.5.}
In Figure~\ref{fig:discussion_model_disagreements}, GPT-5.5 alone matches the non-vulnerable label. The other systems treat unchecked size multiplications as sufficient evidence for an integer-overflow path, whereas GPT-5.5 does not conclude that an out-of-bounds access follows without allocation and downstream-write semantics. The vulnerable function in the same fixing commit is a different function, illustrating a limitation of function-level labels.

Overall, these disagreements often reflect different evidentiary thresholds under incomplete context rather than one model simply ``understanding'' the code better.

\subsection{Explanation Quality Is Not the Same as Downstream Utility}

GPT-5.4 explanations were substantially more accurate than DeepSeek-R1-Qwen-32B explanations in the manual audit, yet the corresponding \SM{} configurations achieve similar downstream performance. Thus, human-rated explanation quality is not the sole determinant of utility. Evidence may be redundant across channels, and \SM{} may learn to ignore unsupported details. Explanation quality should therefore be evaluated intrinsically and through downstream performance.

\subsection{Model Size and Reusability}

A larger \SM{} backbone does not automatically yield substantially better performance. The 1.3B model reaches 72.1\% accuracy, compared with 75.5\% for the 7B model, indicating that scale is only one factor alongside pretraining, fusion design, and explanation representation.

\tool{} caches explanations before \SM{} training, so explanations are generated once per function and can be reused across training epochs, checkpoint evaluations, and subsequent inference on cached functions.

%% file: figures/06_discussion_model_disagreements.tex
\begin{figure*}[t]
    \centering
    \includegraphics[width=0.92\textwidth]{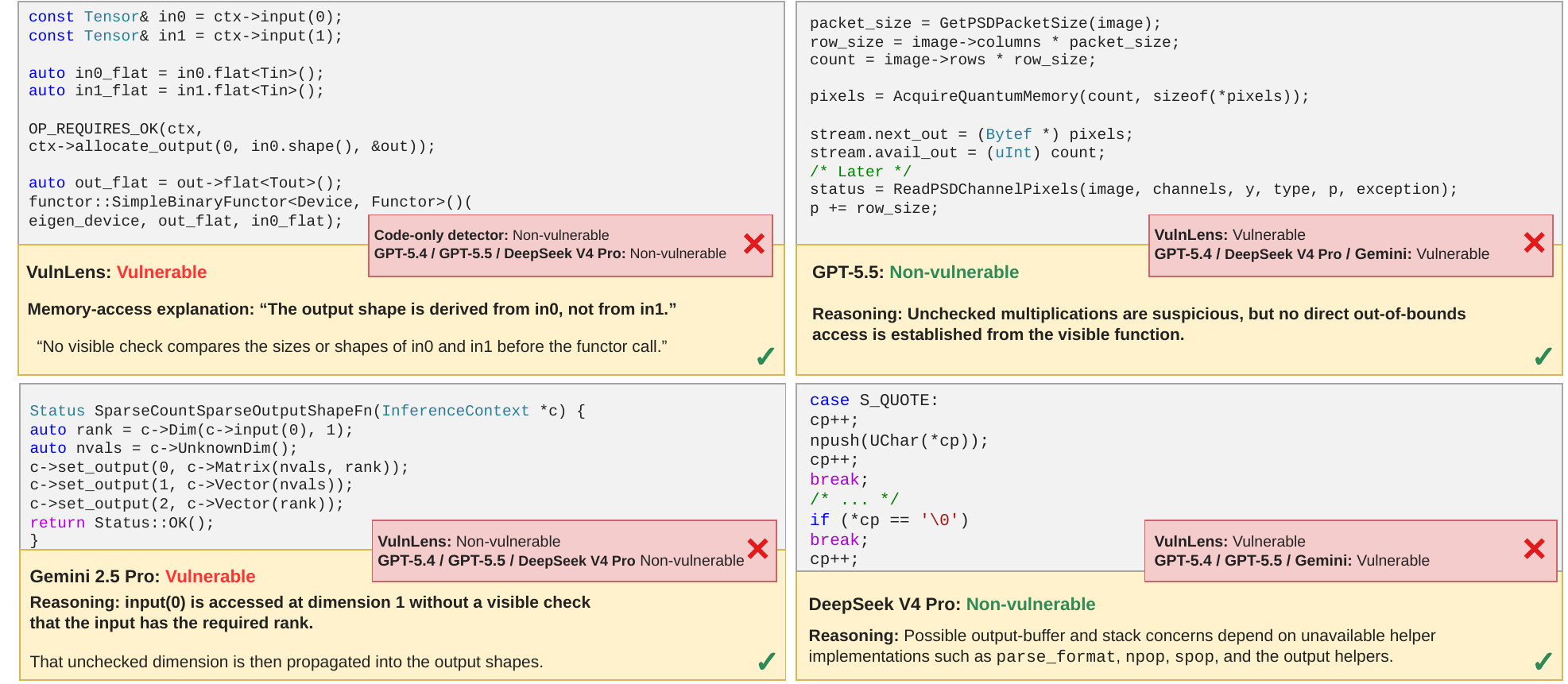}
    \caption{Representative disagreement cases across vulnerability detectors. Each example shows a case where one method uniquely matches the dataset label under incomplete function-level context.}
    \label{fig:discussion_model_disagreements}
\end{figure*}


%% file: sections/07_threats_to_validity.tex
\section{Threats to Validity}
\label{sec:threats}

\noindent\textbf{External validity.}
Our evaluation uses C/C++ functions from PrimeVul and focuses on the ten CWE categories with the largest numbers of functions in that dataset, which may limit generalization to rarer weakness types, other programming languages, projects, or settings requiring broader program context. To reduce dependence on a single benchmark, Section~\ref{sec:rq-transferability} evaluates the same trained \tool{} configuration on SecVulEval without retraining.

\noindent\textbf{Internal validity.}
Threats to internal validity arise from implementation choices, dataset labels, and the reliability of generated explanations. To reduce experimental variation, we use fixed data splits, prompts, hyperparameters, and evaluation procedures when comparing configurations, preserve PrimeVul's original validation and test assignments, and select checkpoints using validation F1 only. Function-level labels and CWE mappings derived from vulnerability-fixing commits may contain noise or depend on surrounding program context. To assess explanation reliability, we manually inspect all seven explanations for 30 stratified functions from each \LLM{} using independent reviewers and report inter-rater agreement.

\noindent\textbf{Construct validity.}
Function-level vulnerability detection may omit caller, callee, type, or project-level context needed to determine whether a function is vulnerable. To characterize performance under this setting, we report accuracy, precision, recall, and F1, while Section~\ref{sec:rq1} evaluates explanation correctness and completeness and Section~\ref{sec:rq6} examines their contribution.

%% file: sections/08_conclusion.tex
\section{Conclusion}
\label{sec:conclusion}

This paper presented \tool{}, an explanation-augmented framework for binary C/C++ vulnerability detection that separates structured program-behavior analysis from the final vulnerability decision. A label-blind \LLM{} analyzes each function from seven fixed perspectives, while a supervised \SM{} integrates the explanations through sequential cross-attention.

On the 526-function PrimeVul test set, \tool{} achieves 75.5\% accuracy and 77.2\% F1, exceeding the strongest evaluated baseline for each metric by 4.2 and 2.2 percentage points, respectively. Compared with a separately trained code-only detector, matched explanations improve F1 by 7.3 points. Without retraining, \tool{} also achieves 66.4\% accuracy and 65.8\% F1 on SecVulEval.

These results show that structured, label-blind explanations can provide complementary predictive information beyond source code alone and serve as effective intermediate representations for vulnerability detection. Future work can extend this approach to richer program context, broader vulnerability categories, and more efficient explanation generation.

%% file: sections/09_Data_Availability.tex
\section{Data Availability}
\label{sec:data_availability}

The replication package for this study is available on
\href{https://zenodo.org/records/23108024?preview=1&token=eyJhbGciOiJIUzUxMiJ9.eyJpZCI6ImZjNzUyMzZiLTVmMzgtNDM0OS04NDFkLTdiNGI3MjI0OGNlZSIsImRhdGEiOnt9LCJyYW5kb20iOiIxYWJhNmZjZjRlMDE3MWQ4ODc2MTU4YTc4Zjk3MzgwOCJ9.Op22-_aqX8Rklm0kJvubyRRNM0lxbZKHWuDO_O8M_t0z34exncHORAwZ8BgHdX4QgHqefRGIMaqWeMbHpB3p9w}{Zenodo (Record 23108024)}.

